# How to Assess the Impact of Quality and Patient Safety Interventions with Routinely Collected Longitudinal Data

Diego A. Martinez, PhD, Assistant Professor of Emergency Medicine, Johns Hopkins University School of Medicine, Baltimore, MD

Mehdi Jalalpour, PhD, Assistant Professor of Civil and Environmental Engineering, Cleveland State University, Cleveland, Ohio, USA.

David T. Efron, MD, FACS, Professor of Surgery, Johns Hopkins University School of Medicine, Baltimore, MD

Scott R. Levin, PhD, Associate Professor of Emergency Medicine, Johns Hopkins University School of Medicine, Baltimore, MD

**ABSTRACT**

**Objectives:** Measuring the effect of patient safety improvement efforts is needed to determine their value but is difficult due to the inherent complexities of hospital operations. In this paper, we show by case study how interrupted time series design can be used to isolate and measure the impact of interventions while accounting for confounders often present in complex health delivery systems.

**Methods:** We searched for time-stamped data from electronic medical records and operating room information systems associated with perioperative patient flow in a large, urban, academic hospital in Baltimore, Maryland. We limited the searched to those adult cases performed between January 2015 and March 2017. We used segmented regression and Box-Jenkins methods to measure the effect of perioperative throughput improvement efforts and account for the loss of high volume surgeons, surgical volume, and occupancy.

**Results:** We identified a significant decline of operating room exit delays of about 50%, achieved in 6 months and sustained over 14 months.

**Conclusions:** By longitudinal assessment of intervention effects, rather than cross-sectional comparison, our measurement tool estimated and provided inferences of change-points over time while taking into account the magnitude of other latent systems factors.

1. INTRODUCTION

Increasingly, interventions are being carried out to improve hospital patient flow and re-design care processes for better quality and patient safety.(1) There is a large body of evidence that links inefficiencies in perioperative throughput to adverse clinical outcomes.(e.g., 2–4) All improvements, however, require change; yet not all change results in improvement. Assessment of the effect of interventions is crucial to identifying valuable change and allocating resources, but it is difficult due to the inherent complexities of healthcare delivery. Furthermore, randomized clinical trials are, most of the time, not conceivable due to impracticality or ethical reasons. In these circumstances, segmented regression of interrupted time series (ITS) designs is a strong, quasi-experimental design tool to estimate the effects of time-delimited interventions in non-randomized settings. This tool allows us to isolate how much an intervention impacted an outcome of interest, immediately and over time; instantly or with delay; transiently or long-term; and, whether factors other than the intervention could explain the change. The goal of this paper is to describe segmented regression (main manuscript) and time series (Supplementary File 2) analysis of ITS data. We use, as of example, improvement efforts steered by the Patient Flow Command Center (PFCC) of the adult perioperative floor of a large, urban, academic hospital in Baltimore, Maryland. This Tool Tutorial includes a brief user's guide in Table 1, an overview of the methods in Sections 2 and 3, and an explanation and example of how to use the tool in Section 4.

## 2. METHODS
### 2.1. Study Setting and Problem Description

In January 2016, the PFCC's leadership voiced the need for measuring the effect of multiple initiatives to streamline patient flow and re-design care processes for improved hospital capacity management. A collaborative composed of engineers and clinicians was formed to develop a systematic approach to analyzing and interpreting the effect of multiple time-delimited interventions. For illustration purposes, we present data here from two interventions to reduce operating room (OR) holds. An OR hold is defined as a delay of over 15 minutes in moving the patient to a postoperative location after the end of a surgical procedure. Following an operative or interventional procedure, especially when given sedation or anesthetic, a patient needs to be medically recovered from their anesthetic event and in most cases, ongoing care administered. For this, specialized nursing care is required, and patients are taken to a post-anesthesia care unit (PACU); once recovered from the anesthetic, the patient may be discharged home (if an outpatient case) or admitted to an appropriate acute care inpatient ward. In the inpatient hospital setting, many patients with the highest postoperative acuity requiring ongoing critical care are transferred to an intensive care unit setting (ICU). In this location, with a higher nursing ratios and skill sets, the patient may also be directly recovered from an anesthetic event. In most acute care hospitals, operating suites have rooms that are used for the entire spectrum of case acuity (from outpatient to critically ill). As a result, postoperative transfer of patients from the OR depends on the ready availability of both PACU and ICU beds. OR holds at the end of a case can result from several causes; however, the primary factor is the ready availability of a patient destination (i.e., downstream bed). Timestamps of the end of the surgical procedure and patient exiting the OR were automatically recorded into the electronic medical records and OR management information system, and thereby the devised interventions, described in the next section, were unlikely to affect our data collection efforts. All adult OR cases over a time span of 114 weeks were included in our analyses. This study was approved by the Johns Hopkins Medicine Institutional Review Board through expedited review.

### 2.2. Interventions to Reduce Operating Room Holds: Central Scheduling & Phase-II PACU

Ready bed availability at the end of an OR case is the primary factor to reduce OR holds. For both ICU and PACU postoperative destinations, the demand for beds is a function of both scheduled and unscheduled cases. While unscheduled cases (emergencies and urgencies) are by definition unpredictable, over time there is a predictable range over which this demand can be calculated. Scheduled cases, on the other hand, can take into account the request for beds on a given day. For example, not all cases on a given day can be scheduled for an ICU bed. Additionally, an increased number of short cases from all rooms might overwhelm a PACU as the exit from the unit may not meet the ongoing demand promptly. Global vision of this system allows for potential smoothing of the demand, at least in the schedulable realm. This is best accomplished by having a *Central Scheduling* mechanism whereby requests and the schedule can be appropriately balanced and optimized within the boundaries of medical necessity. To accomplish this as part of the PFCC, we transitioned to a centralized scheduling model from a process that allowed individual medical office assistants to independently schedule cases regardless of postoperative destination.

Post anesthetic recovery is usually fairly routine and most patients reliable recover on time without incident. When medically stable they are ready to transition to the next level of care or be discharged. In the PACU setting, initial recovery care most basically centers on assurance of vital signs (blood pressure, respiration, and oxygenation), monitoring for postoperative bleeding, as well as assuring pain control and neurologic function. For patients destined for discharge, the ability to tolerate oral rehydration, void, safely ambulate, and accept discharge instructions are subsequent milestones. These latter are noted as _Phase-II recovery_ and require less intensive nursing (with lower nurse/patient ratios). Regardless of whether a patient is destined to be admitted or discharged, post-anesthetic recovery represents a patient occupying a position in the PACU. Many outpatient surgical PACUs will cohort patients with similar intensities for efficiency. That is patients who are ready for Phase II may move to an area with other patients, be maintained in a reclining chair (as opposed to a stretcher) to facilitate mobility, and per oral challenge, which also allows reducing nurse to patient ratios. Up to 50% of cases performed at one of our inpatient operating suites were outpatient cases. The adjoining PACU was designed primarily for inpatient flow. To optimize the flow of outpatient cases in the inpatient suite setting, an unutilized additional PACU area (adjacent to another operating suite) was opened as a _Phase-II PACU_ area and staffed at an appropriate nursing ratio. Outpatients who are physiologically clear are then transferred to the Phase-II location, thereby opening a slot for a fresh PACU admission.

Important to note is that supply of beds in the postoperative destinations (PACU and ICU) is a function of physical location and staffing (open and staffed beds) and successful discharge of patients to the next level of care. Exits of patients from the PACU is dependent on primarily two parameters: 1) recovery time from the anesthetic (both for inpatient and outpatient cases) and 2) an available inpatient acute care bed. Exits from an ICU depend primarily on the latter, and at times, the need to transfer patients from both the PACU and ICU will compete for the same downstream resource. Discharges from the hospital often occur later in the day, adding inherent delays to timely physical bed space available for the patient's final inpatient destination. As such, fluctuations in bed occupancy and patient volume in surgical floor beds are hypothesized to be significant confounding factors in the assessment.

**Table 1.** Segmented Regression of Interrupted Time Series in Quality and Patient Safety Studies (User's Guide)

**Problem Statement:** Measurement of the impact of quality and patient safety interventions is critical to identifying valuable change and allocating scarce resources, but it is difficult due to common non-randomized settings and the inherent complexities of patient flow and healthcare delivery.

**The Purpose of the Tool:** The purpose of this tool is to provide a systems approach to help care givers and administrators measure changes in quality and patient safety metrics while controlling for secular trends that may have occurred without the intervention. The three broad components of measurement are determining key metrics, collecting an appropriate amount of data, and analyzing and interpreting the data; this Tool Tutorial focuses on the last component.

**Who Should Use This Tool:** The tool should be used by analysts and process improvement champions seeking to measure the impact of quality and safety improvement interventions. Clinical and operational staff involved in changes to healthcare processes should be included in the discussion of identifying potential confounder factors that may bias the effect of the interventions.

**How To Use This Tool:** The tool is best applied in the context of comprehensive quality and safety improvement efforts, such as hospital capacity command centers across the country. This tool should be used periodically to monitor progress and sustainability of improvement efforts.

### 2.3. Tool Description and How To

A _time series_ is a sequence of values of a particular measure taken at regularly spaced intervals over time. _Segments_ in a time series are defined when the sequence of measures is divided into two or more portions at change points. _Change points_ are specific points in time where the values of the series may exhibit a change from the previously established pattern because of an identifiable real-world event, a policy change, or a new process improvement intervention. The choice of the beginning and end of the intervention, with the possible addition of some pre-specified lag time to allow the intervention to take effect. _Segmented regression_ is a method for statistical modeling the ITS data to draw isolated inferences about the effect of a real-world event, policy, or intervention on the measure of interest.

Two parameters describe each segment of a time series: level and trend. The _level_ is the value of the series at the beginning of a given time interval (i.e., the y-intercept for the first segment and the value immediately following each change point at which successive segments join). The _trend_ is the rate of change of a measure (i.e., the slope) during a segment. Particularly, in segmented regression, each segment of the series is allowed to exhibit both a level and a trend. The analyst examines the changes in level and trend that follows an intervention. A change in level, i.e., a jump or drop in the outcome after the intervention, constitutes an abrupt intervention effect. A change in trend is defined by an increase or decrease in the slope of the segment after the intervention as compared with the segment preceding the intervention. A change in trend represents a gradual change in the value of the outcome during the segment. Segmented regression uses statistical models to estimate the level and trend in the pre-intervention segment, and then calculates changes in level and trend after the intervention.

Figure 1 shows the time series of the total number of OR holds per week in the surgery department in a cohort of 18,838 patients, with an average of 23 OR holds per week. Beginning in January 2016 (week 53), the PFCC implemented a process improvement program including both Central Scheduling and Phase-II PACU. The improvement program interrupts the time series in two segments of interest. Figure 1 shows an abrupt level change in the total number of OR holds, from about 30 per week to 15 per week, followed the improvement program. There was a very little month-to-month change in the number of OR holds before as well as after the intervention.

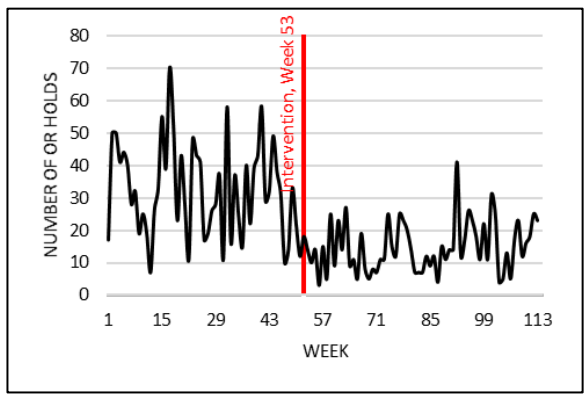

**Figure 1.** Time series of the total number of operating room holds (i.e., OR exit delays) in the surgery service.

### 2.4. Data Sources and Measures

Segmented regression requires data collected regularly over time, and organized at equally spaced intervals (for ease of modeling). Routinely maintained patient flow data and electronic medical records, as well as cost data, are commonly used as sources of time series data for quality and patient safety studies. Although these data sources may not have been gathered for research purposes, they often provide reliable measures of relevant dependent variables for ITS studies.

Outcome measures for longitudinal studies can include the use of healthcare resources and clinical measures. Outcomes can be expressed in averages, proportions, or rates. Examples of quality and patient safety measures are the average number of diagnostic errors per patient or the number of near-misses per week. Examples of utilization measures would be the average length of hospital stay and the monthly rate of admission to nursing homes. Finally, examples of clinical measures might include average heart rate, the percentage of diabetic patients achieving adequate glucose control, or mortality rates.

A sufficient number of time points before and after the intervention is needed to conduct ITS analysis. A general recommendation is for 18 data points before and 18 data points after the intervention, although the analyst should be cautious when conducting studies with a sample size as small as 18 time points per segment.(5) Rather, with 24 monthly measures, the analyst can adequately evaluate seasonal variation. Also, there needs to be a sufficient number of observations at each data point; some may say a minimum of 100 is desirable. However, there are important statistical and ethical implications in the choice of sample size for a study and proper techniques should be used to estimate a sufficient number of observations. We direct the reader to (6) to learn more about methods to estimate sample size.

3. RESULTS AND DISCUSSION
    3.1. Statistical Analysis and Interpretation of the Results

A strength of segmented regression is the intuitive graphical representation of the results. Visually, we compare the time series pattern before the intervention with the pattern after the intervention and assess if, after the intervention, the time series pattern has changed in relation to the pre-intervention pattern. Looking at the data points in Figure 1, we would have expected the pre-intervention series to continue at an average of about 32 OR holds per week had the improvement program not occurred. Clearly, after the interventions, the average number of OR holds was about half of what would have been expected. Although we can often detect changes in level and trend by looking at a time series, we cannot easily see whether changes in level and trend could be the result of secular trends, factors other than the intervention, or chance alone. To measure change and control for secular trends and other confounders, segmented regression analysis of ITS data is illustrated and applied next.

Common segmented regression models fit a least squared regression line to each segment of the independent variable, time, and thus assume a linear relationship between time and the outcome within each segment. We can specify the following linear regression model to estimate the level and trend in OR hold volume before the intervention and the changes in level and trend following the intervention:

$$y_t = \beta_0 + \beta_1 * time_t + \beta_2 * intervention_t + \beta_3 * time\ after\ intervention_t + \boldsymbol{X}^T \boldsymbol{\beta_X} + \epsilon_t \quad (1)$$

In (1), $y_t$ is the observed outcome variable in week $t$ (OR hold volume); $time_t$ is a numeric variable indicating time in weeks at time $t$ from the start of the observation period; $intervention_t$ is an indicator for time $t$ occurring before ($intervention_t = 0$) or after ($intervention_t = 1$) the intervention, which was implemented at week 53 in the series shown in Figure 1; and $time\ after\ intervention_t$ is a numeric variable counting the number of weeks after the intervention at time $t$, coded as 0 before the intervention and ($time$-53) after the intervention. In this model, $\beta_0$ estimates the baseline level of the outcome variable (OR holds) at time zero; $\beta_1$ estimates the week-to-week change of the outcome variable before the intervention (i.e., baseline trend); $\beta_2$ estimates the level change of the outcome immediately after the intervention; $\beta_3$ estimates the change in trend in the outcome after the intervention as compared to the weekly trend before the intervention; $\beta_X$ estimates the effect of other concurrent confounders, e.g., loss of high volume surgeons, OR case volume, and bed occupancy; and, the error term $\epsilon_t$ at time $t$ represents random variability not explained by the model. The error term consists of a normally distributed random error at time $t$ that may be correlated to errors at preceding or subsequent time points (autocorrelation). Autocorrelation of the errors violates the least squared regression assumption that the error terms are uncorrelated, resulting in underestimation of the standard errors and overestimation of the p-values. Using (1) to estimate the level and trend changes allows us to control for baseline level, trend, and potential confounders. Using (SF1, see Supplementary File) allows us to control for baseline level, trend, potential confounders, and autocorrelation—a major strength of segmented regression and time series analysis commonly encountered in quality and patient safety evaluations.

We wanted to isolate and evaluate statistically the effect of the intervention amongst these other confounders that we hypothesized would influence OR holds. Table 2 contains the parameter estimates from the segmented regression model predicting weekly OR hold volume. Using the results in Table 2 and knowing that there were 189 admissions, 169 discharges, and average bed occupancy reached 81.1% (see Supplementary File 1), we estimated that in week 54 the volume of OR holds was 11. Had the interventions not being introduced, the number of OR holds would have been 28. Thus, the number of OR holds per week decreased by 17, or 60% (95% confidence interval -54%, -70%) after the improvement program was implemented, compared with what it would have been without the program.

**Table 2.** Parameter estimates from the full and most parsimonious segmented regression models predicting operating room holds (i.e., OR exit delays) in the surgery service.

|  | Coefficient | Standard Error | T-Statistic | P-Value | |
|---|---|---|---|---|---|
| Intercept $\beta_0$ | -50.64 | 19.10 | -2.66 | 0.009 | ** |
| Baseline trend $\beta_1$ | -0.10 | 0.10 | -1.08 | 0.283 | |
| Level change $\beta_2$ | -12.01 | 4.05 | -2.97 | 0.004 | ** |
| Trend change $\beta_3$ | 0.16 | 0.12 | 1.32 | 0.190 | |
| Admissions $\beta_4$ | 0.07 | 0.07 | 1.15 | 0.250 | |
| Discharges $\beta_5$ | -0.11 | 0.08 | -1.26 | 0.210 | |
| Occupancy $\beta_6$ | 1.04 | 0.22 | 4.78 | <0.001 | *** |

Significance level: * ≤0.1; ** ≤0.05; *** ≤0.01. Abbreviations: SE, standard error; ED, emergency department; OR, operating room.

### 3.2. Data structure for the analysis

Supplementary File 1 illustrates the data structure of the previous analysis. The aggregated outcome measures, average ED boarding duration and the total number of OR holds, are calculated at each weekly time point. We specified a pre-intervention trend variable and a level and trend change variables for each of the time series.

Different definitions of the time variable in segmented regression analysis are possible. For example, time could be rescaled so that the starting point of the intervention is coded as week 1, with time being measured backward and forward from that point. Alternatively, time at the point of interest, for example, 53 weeks after the intervention, could be coded as 1, with time counted backward and forward from there. Recording time in these ways only changes the interpretation of the intercept. It does not change the absolute or relative measures of effect.

### 3.3. Correcting for autocorrelation and other explanatory factors

Ordinary least squares regression analysis assumes that error terms associated with each observation are uncorrelated. That is the differences between the actual outcome value and those predicted by the regression model are uncorrelated. As time is incorporated as a predictor in the segmented regression models, error terms of consecutive observations are often correlated. That is, hospital operations metrics at two time points that are close to each other are more similar than performance metrics at two time points further apart, resulting in autocorrelated errors. Failing to correct for autocorrelation may lead to underestimated standard errors and overestimated significance of the effects of interventions. Fortunately, one can detect autocorrelation and available statistical modeling and software con control for it.

One can detect autocorrelation by inspecting plot of residuals against time and conducting statistical tests. Randomly scattered residuals indicate there is no autocorrelation; otherwise, there is an indication that autocorrelation is present. The Durbin-Watson (DW) statistic can be used to test for autocorrelation of the residuals.(7) In our example, the Durbin-Watson statistic tested negative for autocorrelation of the residuals (Model A, DW statistic 1.79 and P-Value=0.144; Model B, DW statistic 1.66 and P-Value=0.076). However, another type of autocorrelation was detected. That is autocorrelation of the outcome variable, number of OR holds per week. If we want to take advantage of such autocorrelation to reach a better fit between the model and actual values, a second approach, time series analysis, is described in the Supplementary File 2.

The effect of interventions aiming at improving hospital operations are influenced by a number of interrelated patient flow metrics. In our context, the volume of OR holds is associated with fluctuations of bed occupancy and inpatient throughput (as described in Section 2.1). During high occupancy or high patient volume weeks, it is expected to have more OR holds due to unavailability of beds to place patients after the surgical procedure is finished. To account for these system-level factors and their confounding effects, we incorporated the average hourly bed occupancy per week, the number of admissions per week, and the number of discharges per week into the model as predictor variables.

## 4. SUMMARY AND NEXT STEPS

This paper illustrates general points about how to use segmented regression of ITS in process improvement efforts at hospitals. Segmented regression can be used in the daily management of hospital operations to analyze routinely collected data and promote evidence-based management. Segmented regression can also help hospital leadership avoid fruitless investments in changes that sound promising but do not deliver results. For the OR holds improvement program, for example, the results have been substantial: a decline of OR exit delays of about 60%, achieved in 6 months and sustained over 14 months while taking into account factors other than the intervention that could have explained the improvement (e.g., fluctuations in patient volume and occupancy).

Segmented regression has important limitations. First, the models presented here assume a linear trend in the outcome within each segment (pre- and post-intervention). Changes may follow non-linear patterns; for instance, intervention effects may have an increasing or decreasing curvilinear trend. The non-linear patters may require other modeling approaches including time series (Box-Jenkins) modeling.[see Supplementary File 2] Although these models are widely used for forecasting future trends, they are of less use in understanding changes in trend that occur at defined time points and impose further complexity to communicate results. Second, segmented regression typically aggregates individual-level data by time point (day, week, month, and year). Contrary to cross-sectional approaches such as linear or logistic regression, segmented regression of ITS data does not allow control for individual-level covariates. Individual-characteristics, however, would only confound the time series results if they predicted the outcome and changed the relationship to the time of the intervention.

To improve hospital operations, we must measure the effect of process improvement interventions. Segmented regression is a robust modeling approach that allows the analyst to estimate the effect of time-delimited interventions while controlling for secular changes that may have occurred in the absence of the intervention. In the hands of hospital operations specialists, segmented regression can enable reflection and assisting in the identification of further improvement opportunities.

## 5. COMPLIANCE WITH ETHICAL STANDARDS

The authors whose names are listed above certify that they have no affiliations with or involvement in any organization or entity with any financial interest, or non-financial interest in the subject matter or materials discussed in this manuscript.

**SUPPLEMENTARY FILE 1**

**Data Structure for the Analysis**

The structure of data for analysis of the impact of multiple interventions on the number of operating room holds (i.e., OR exit delays) per week. OR hold is defined as any patient waiting in the OR post-procedure for an intensive care unit or a post-anesthesia care unit for more than 15 minutes.

| Week | OR Holds Volume | Surgery Service Occupancy | Surgery Service Discharges | Surgery Service Admissions | Level Change | Trend Change | Baseline Trend |
|---|---|---|---|---|---|---|---|
| 1 | 16 | 65.5 | 156 | 212 | 0 | 0 | 1 |
| 2 | 17 | 77.4 | 156 | 212 | 0 | 0 | 2 |
| 3 | 50 | 85.8 | 183 | 232 | 0 | 0 | 3 |
| 4 | 50 | 82.4 | 164 | 184 | 0 | 0 | 4 |
| 5 | 41 | 85 | 188 | 200 | 0 | 0 | 5 |
| 6 | 44 | 86.8 | 172 | 218 | 0 | 0 | 6 |
| 7 | 40 | 86.7 | 177 | 201 | 0 | 0 | 7 |
| 8 | 28 | 86.9 | 173 | 195 | 0 | 0 | 8 |
| 9 | 32 | 85.9 | 174 | 210 | 0 | 0 | 9 |
| 10 | 19 | 84.3 | 199 | 208 | 0 | 0 | 10 |
| 11 | 25 | 86 | 169 | 215 | 0 | 0 | 11 |
| 12 | 19 | 84.8 | 179 | 213 | 0 | 0 | 12 |
| 13 | 7 | 80.5 | 179 | 186 | 0 | 0 | 13 |
| 14 | 26 | 85.3 | 179 | 222 | 0 | 0 | 14 |
| 15 | 33 | 88.6 | 160 | 197 | 0 | 0 | 15 |
| 16 | 55 | 89.7 | 188 | 192 | 0 | 0 | 16 |
| 17 | 39 | 89 | 167 | 194 | 0 | 0 | 17 |
| 18 | 70 | 89.9 | 162 | 188 | 0 | 0 | 18 |
| 19 | 52 | 91.8 | 164 | 173 | 0 | 0 | 19 |
| 20 | 23 | 90 | 164 | 188 | 0 | 0 | 20 |
| 21 | 43 | 87.9 | 178 | 195 | 0 | 0 | 21 |
| 22 | 27 | 81 | 180 | 186 | 0 | 0 | 22 |
| 23 | 11 | 80.5 | 176 | 194 | 0 | 0 | 23 |
| 24 | 48 | 85.1 | 158 | 194 | 0 | 0 | 24 |
| 25 | 43 | 90.2 | 164 | 207 | 0 | 0 | 25 |
| 26 | 41 | 88.7 | 186 | 186 | 0 | 0 | 26 |
| 27 | 17 | 87.9 | 178 | 190 | 0 | 0 | 27 |
| 28 | 19 | 84.5 | 183 | 185 | 0 | 0 | 28 |
| 29 | 26 | 87 | 189 | 207 | 0 | 0 | 29 |
| 30 | 28 | 87.3 | 175 | 212 | 0 | 0 | 30 |
| 31 | 37 | 88.8 | 178 | 197 | 0 | 0 | 31 |
| 32 | 11 | 85.7 | 158 | 188 | 0 | 0 | 32 |
| 33 | 58 | 84.6 | 189 | 203 | 0 | 0 | 33 |
| 34 | 16 | 82.3 | 182 | 204 | 0 | 0 | 34 |
| 35 | 37 | 84.7 | 176 | 212 | 0 | 0 | 35 |
| 36 | 24 | 85.1 | 176 | 195 | 0 | 0 | 36 |
| 37 | 15 | 81.1 | 155 | 178 | 0 | 0 | 37 |
| 38 | 40 | 85.3 | 174 | 196 | 0 | 0 | 38 |
| 39 | 22 | 87.6 | 175 | 209 | 0 | 0 | 39 |
| 40 | 39 | 92 | 148 | 166 | 0 | 0 | 40 |
| 41 | 43 | 88.3 | 183 | 204 | 0 | 0 | 41 |
| 42 | 58 | 89.7 | 158 | 185 | 0 | 0 | 42 |
| 43 | 29 | 85.9 | 172 | 186 | 0 | 0 | 43 |
| 44 | 32 | 86.8 | 162 | 201 | 0 | 0 | 44 |
| 45 | 49 | 88.7 | 179 | 199 | 0 | 0 | 45 |
| 46 | 38 | 88 | 177 | 219 | 0 | 0 | 46 |
| 47 | 31 | 88 | 185 | 212 | 0 | 0 | 47 |
| 48 | 10 | 75 | 141 | 126 | 0 | 0 | 48 |
| 49 | 14 | 82.7 | 161 | 198 | 0 | 0 | 49 |
| 50 | 33 | 85.5 | 168 | 199 | 0 | 0 | 50 |
| 51 | 22 | 87 | 182 | 197 | 0 | 0 | 51 |
| 52 | 12 | 68.2 | 184 | 136 | 0 | 0 | 52 |
| 53 | 18 | 62.8 | 135 | 189 | 1 | 1 | 53 |
| 54 | 14 | 81.8 | 169 | 189 | 1 | 2 | 54 |
| 55 | 10 | 85.7 | 141 | 153 | 1 | 3 | 55 |
| 56 | 14 | 78.5 | 173 | 159 | 1 | 4 | 56 |
| 57 | 3 | 79.7 | 153 | 176 | 1 | 5 | 57 |
| 58 | 15 | 78.3 | 144 | 179 | 1 | 6 | 58 |
| 59 | 5 | 81.4 | 163 | 189 | 1 | 7 | 59 |
| 60 | 25 | 81.6 | 164 | 196 | 1 | 8 | 60 |
| 61 | 9 | 83.9 | 158 | 183 | 1 | 9 | 61 |
| 62 | 23 | 79.3 | 170 | 205 | 1 | 10 | 62 |
| 63 | 14 | 84 | 164 | 182 | 1 | 11 | 63 |
| 64 | 27 | 86 | 182 | 189 | 1 | 12 | 64 |
| 65 | 9 | 85.9 | 170 | 171 | 1 | 13 | 65 |
| 66 | 11 | 78.7 | 145 | 179 | 1 | 14 | 66 |
| 67 | 5 | 81.5 | 164 | 186 | 1 | 15 | 67 |
| 68 | 19 | 79.5 | 163 | 207 | 1 | 16 | 68 |
| 69 | 8 | 83.4 | 175 | 190 | 1 | 17 | 69 |
| 70 | 5 | 86.3 | 172 | 173 | 1 | 18 | 70 |
| 71 | 8 | 82.3 | 157 | 184 | 1 | 19 | 71 |
| 72 | 7 | 82.6 | 163 | 177 | 1 | 20 | 72 |
| 73 | 11 | 78.8 | 152 | 181 | 1 | 21 | 73 |
| 74 | 11 | 85.7 | 139 | 154 | 1 | 22 | 74 |
| 75 | 25 | 81 | 187 | 188 | 1 | 23 | 75 |
| 76 | 15 | 79.5 | 156 | 168 | 1 | 24 | 76 |
| 77 | 12 | 76.7 | 156 | 182 | 1 | 25 | 77 |
| 78 | 25 | 75.8 | 167 | 191 | 1 | 26 | 78 |
| 79 | 23 | 73.9 | 109 | 134 | 1 | 27 | 79 |
| 80 | 20 | 84.6 | 164 | 180 | 1 | 28 | 80 |
| 81 | 14 | 82.7 | 144 | 163 | 1 | 29 | 81 |
| 82 | 7 | 76.2 | 165 | 161 | 1 | 30 | 82 |
| 83 | 7 | 73.3 | 150 | 179 | 1 | 31 | 83 |
| 84 | 7 | 79.5 | 157 | 173 | 1 | 32 | 84 |
| 85 | 12 | 83.3 | 184 | 178 | 1 | 33 | 85 |
| 86 | 9 | 78.5 | 165 | 184 | 1 | 34 | 86 |
| 87 | 12 | 79.7 | 161 | 195 | 1 | 35 | 87 |
| 88 | 4 | 77.3 | 150 | 180 | 1 | 36 | 88 |
| 89 | 15 | 83.8 | 172 | 204 | 1 | 37 | 89 |
| 90 | 11 | 85.6 | 166 | 160 | 1 | 38 | 90 |
| 91 | 14 | 84.4 | 162 | 184 | 1 | 39 | 91 |
| 92 | 14 | 87.2 | 164 | 185 | 1 | 40 | 92 |
| 93 | 41 | 85.7 | 148 | 175 | 1 | 41 | 93 |
| 94 | 12 | 84.2 | 167 | 186 | 1 | 42 | 94 |
| 95 | 17 | 84.1 | 174 | 193 | 1 | 43 | 95 |
| 96 | 26 | 85.9 | 164 | 169 | 1 | 44 | 96 |
| 97 | 23 | 85.8 | 157 | 176 | 1 | 45 | 97 |
| 98 | 18 | 86.2 | 174 | 182 | 1 | 46 | 98 |
| 99 | 11 | 79.2 | 168 | 151 | 1 | 47 | 99 |
| 100 | 22 | 80.8 | 150 | 206 | 1 | 48 | 100 |
| 101 | 11 | 85.7 | 170 | 190 | 1 | 49 | 101 |
| 102 | 31 | 85.5 | 178 | 195 | 1 | 50 | 102 |
| 103 | 25 | 84.1 | 176 | 190 | 1 | 51 | 103 |
| 104 | 4 | 67 | 125 | 119 | 1 | 52 | 104 |
| 105 | 5 | 71.7 | 123 | 151 | 1 | 53 | 105 |
| 106 | 13 | 78.6 | 174 | 193 | 1 | 54 | 106 |
| 107 | 5 | 73.6 | 127 | 166 | 1 | 55 | 107 |
| 108 | 17 | 83.9 | 171 | 202 | 1 | 56 | 108 |
| 109 | 23 | 84.7 | 146 | 165 | 1 | 57 | 109 |
| 110 | 12 | 81.3 | 162 | 181 | 1 | 58 | 110 |
| 111 | 16 | 82.8 | 153 | 170 | 1 | 59 | 111 |
| 112 | 18 | 86.2 | 160 | 186 | 1 | 60 | 112 |
| 113 | 25 | 81.6 | 152 | 183 | 1 | 61 | 113 |
| 114 | 23 | 85.4 | 141 | 170 | 1 | 62 | 114 |

**SUPPLEMENTARY FILE 2**

**Intervention Analysis with Time Series Regression**

To account for effects of serial autocorrelation between the outcomes and impacts of exogenous variables on the outcome, we use time series models. The class of time series models we use are autoregressive models with exogenous variables $ARX(\rho)$, where $p$ is the autoregressive order. These models are expressed as follows:

$$y_t = x^T \beta + \sum_j^p \phi_j(y_{t-j} - x_{t-j}\beta) + \epsilon_t, \quad \text{(SF1)}$$

where $y_t$ is the response (outcome variable) at time $t$; $x$ is a vector that collects the exogenous variables (including intervention variables); and, $\epsilon_t$ are the residuals which are assumed to be normally distributed, zero-mean and uncorrelated with constant variance. Model parameters are $\beta$ and $\phi$, which are estimated from the data using maximum likelihood method. To evaluate the effect of intervention on the outcome data, we pursue an approach similar to (5), where this evaluation is translated into determining whether adding the intervention exogenous variables to the time series model leads to a statistically significant change in model goodness of fit. Goodness of fit for each model is measured with model deviance, which is defined as twice the negative of log likelihood function magnitude (alternatively, Akaike or Bayesian Information Criteria could be used). Deviance allows using a likelihood ratio test for model selection as follows. We begin with estimating the best "baseline" model, which includes appropriate autoregressive order and all statistically significant exogenous variables other than the intervention variables, and evaluate its deviance with the data. The selected model has the smallest deviance and should not have autocorrelated residuals (6–8). Next, intervention variables are added to this model, yielding the "full model", and then the parameters of the full model are estimated using the maximum likelihood method. Note that the baseline model is nested within the full model. Next, the difference in deviances defined as:

$$\Lambda = D_b - D_f, \quad \text{(SF2)}$$

where $D_b$ is the deviance for the baseline model and $D_f$ is the deviance for the full model, is used for the statistical test. From (7,8), it follows that $\Lambda$ follows $\chi_\nu^2$ distribution with $\nu$ degrees of freedom. The degrees of freedom is equal to the difference of number of parameters between these two models. For example, if both the intervention trend and the intervention level change are added to the full model, then $\nu = 2$. If $\Lambda$ is greater than the critical values of the $\chi^2$ distribution at the selected significance level, for instance $\alpha = 0.05$, then intervention variables are significant and it is inferred that the intervention has a positive impact on the outcome data. A key component of the proposed methodology is to estimate the best baseline model (i.e., the model with minimum $D_b$) using all the other information other than the intervention variables to clearly isolate the intervention effects. We illustrate how this statistical approach is used with the OR holds volume example presented in the main manuscript.

Following the procedure outlined in (9), we determined the best baseline model with $ARX(2)$ and intercept and Occupancy rates, which resulted in $D_b = 847.31$. Next, we added intervention level change variable and estimated the full model with $D_f = 835.15$, which results in $\Lambda = 12.18 > 3.84$. We also added the trend change, which resulted in a worse fit. Therefore, it is inferred that the intervention is statistically significant to changing the level, but not the trend of OR holds. Note that both models pass the residual check, which implies they are likely to be adequate. We note that the deviance for the best segmented regression model was 852.84 (includes only the variables in Part B of Table SF1 without autoregressive terms), which highlights how adding the autoregressive terms results in a significantly better fit. However, in the particular example of weekly OR holds, the inferences matched with those of segmented regression models. Predictions from these models are shown in Figure SF1, where a lower level in the full model predictions are observed after the intervention. We note that because of the Gaussian assumption in time series, and that low counts are being predicted, the prediction is negative at one instance. This could be remedied if generalized autoregressive models as presented in (7,9,10) are used.

**Table SF 1.** Parameter estimates from the full segmented regression and most parsimonious time series models predicting operating room exit delays (OR holds) in the surgery service.

|  | Mag | SE |
|---|---:|---:|
| A. Segmented regression model | | |
| Intercept $\beta_0$ | -52.57 | 20.57 |
| Baseline trend $\beta_1$ | -0.11 | 0.12 |
| Level change $\beta_2$ | -11.87 | 4.12 |
| Trend change $\beta_3$ | 0.17 | 0.14 |
| Admissions $\beta_4$ | 0.07 | 0.07 |
| Discharges $\beta_5$ | -0.10 | 0.08 |
| Occupancy $\beta_6$ | 1.00 | 0.24 |
| B. Most parsimonious time series model | | |
| Intercept $\beta_0$ | -55.48 | 18.63 |
| Autoregressive $\phi_1$ | 0.03 | 0.09 |
| Autoregressive $\phi_2$ | 0.19 | 0.09 |
| Level change $\beta_2$ | -12.59 | 2.73 |
| Occupancy $\beta_6$ | 1.02 | 0.21 |

**Figure SF1.** Time series models predicting the number of OR holds per week. Baseline model incorporates the intervention. Full model incorporates the intervention and the potential confounders.

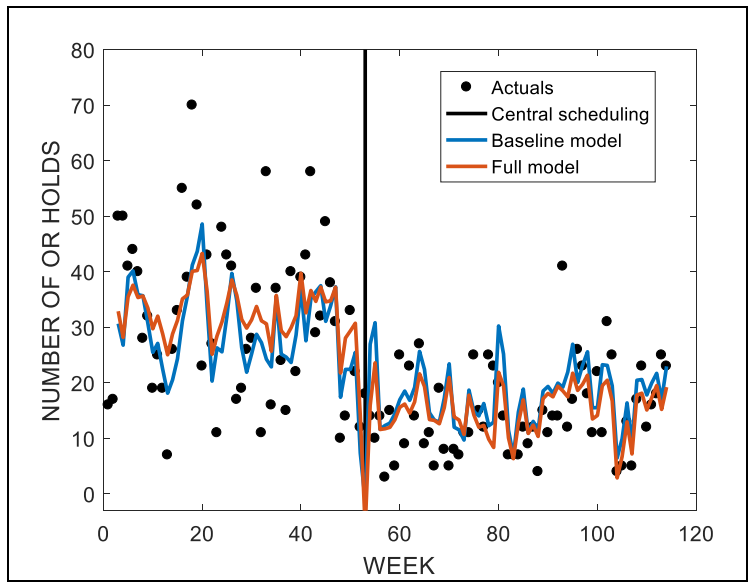